\documentclass[]{spie}  

\usepackage{amsmath,amsfonts,amssymb}
\usepackage{graphicx}
\usepackage[colorlinks=true, allcolors=blue]{hyperref}
\usepackage{color}

\title{Review of high-contrast imaging systems for current and future ground-based and space-based telescopes III.\\ Technology opportunities and pathways}

\author[a]{Frans Snik}
\author[b]{Olivier Absil}
\author[c]{Pierre Baudoz}
\author[d]{Mathilde Beaulieu}
\author[e]{Eduardo Bendek}
\author[f]{Eric Cady}
\author[b]{Brunella Carlomagno}
\author[g]{Alexis Carlotti}
\author[c]{Nick Cvetojevic}
\author[a]{David Doelman}
\author[h]{Kevin Fogarty}
\author[c]{\\Rapha\"{e}l Galicher}
\author[i,j,k]{Olivier Guyon}
\author[a]{Sebastiaan Haffert}
\author[c]{Elsa Huby}
\author[f]{Jeffrey Jewell}
\author[l]{Nemanja Jovanovic}
\author[a]{Christoph Keller}
\author[a]{Matthew A. Kenworthy}
\author[j,m]{Justin Knight}
\author[n]{Jonas K\"uhn}
\author[h,o]{Johan Mazoyer}
\author[j,m]{Kelsey Miller}
\author[d]{Mamadou N'Diaye}
\author[p,q]{Barnaby Norris}
\author[a]{Emiel Por}
\author[h]{Laurent Pueyo}
\author[f]{A J Eldorado Riggs}
\author[l]{Garreth Ruane}
\author[e]{Dan Sirbu}
\author[f]{J. Kent Wallace}
\author[a]{Michael Wilby}
\author[r]{Marie Ygouf}
\affil[a]{Leiden Observatory, Leiden University, P.O. Box 9513, 2300 RA Leiden, The Netherlands}
\affil[b]{Space sciences, Technologies, and Astrophysics Research (STAR) Institute, University of Li\`ege, 19C all\'ee du Six Ao\^ut, B-4000 Li\`ege, Belgium}
\affil[c]{LESIA, Observatoire de Paris, PSL Research University, CNRS, Sorbonne Universit\'es, Univ. Paris Diderot, UPMC Univ. Paris 06, Sorbonne Paris Cit\'e, 5 place Jules Janssen, 92190 Meudon, France}
\affil[d]{Universit\'e Cote d'Azur, Observatoire de la Cote d'Azur, CNRS, Laboratoire Lagrange, Parc Valrose, F-06108 Nice, France}
\affil[e]{NASA Ames Research Center, Moffett Field, Mountain View, CA, 94035, USA}
\affil[f]{Jet Propulsion Laboratory, California Institute of Technology, 4800 Oak Grove Drive, Pasadena, CA 91109 USA}
\affil[g]{CNRS, Institut de Plan\'etologie et d'Astrophysique de Grenoble (IPAG), F-38000, Grenoble, France}
\affil[h]{Space Telescope Science Institute, 3700 San Martin Drive, 21218 Baltimore MD, USA}
\affil[i]{Astrobiology Center, National Institutes of Natural Sciences, 2-21-1 Osawa, Mitaka, Tokyo, JAPAN}
\affil[j]{Steward Observatory, University of Arizona, Tucson, AZ 85721, USA}
\affil[k]{National Astronomical Observatory of Japan, Subaru Telescope, National Institutes of Natural Sciences, Hilo, HI 96720, USA}
\affil[l]{California Institute of Technology, 1200 E. California Blvd., Pasadena, CA 91125, USA}
\affil[m]{College of Optical Sciences, University of Arizona, 1630 E University Blvd, Tucson, AZ 85719, USA}
\affil[n]{Institute for Particle Physics and Astrophysics, ETH Zurich, Wolfgang-Pauli-Str. 27, CH-8093 Zurich, Switzerland}
\affil[o]{Johns Hopkins University, Zanvyl Krieger School of Arts and Sciences, Department of Physics and Astronomy, Bloomberg Center for Physics and Astronomy, 3400 North Charles Street, Baltimore, MD 21218, USA}
\affil[p]{Sydney Institute for Astronomy (SIfA), School of Physics, University of Sydney, NSW 2006, Australia}
\affil[q]{Australian Astronomical Observatory, Faculty of Science and Engineering, Macquarie University, NSW 2109, Australia}
\affil[r]{IPAC, Caltech, 1200 E. California Blvd., Pasadena, CA 91125, USA}

\authorinfo{Further author information: (Send correspondence to Frans Snik)\\Frans Snik: E-mail:  snik@strw.leidenuniv.nl}

\begin{document} 
\maketitle

\begin{abstract}
The Optimal Optical Coronagraph Workshop at the Lorentz Center in September 2017 in Leiden, the Netherlands gathered a diverse group of 25 researchers working on exoplanet instrumentation to stimulate the emergence and sharing of new ideas. 
This contribution is the final part of a series of three papers summarizing the outcomes of the workshop, and presents an overview of novel optical technologies and systems that are implemented or considered for high-contrast imaging instruments on both ground-based and space telescopes. 
The overall objective of high contrast instruments is to provide direct observations and characterizations of exoplanets at contrast levels as extreme as $10^{-10}$. 
We list shortcomings of current technologies, and identify opportunities and development paths for new technologies that enable quantum leaps in performance. 
Specifically, we discuss the design and manufacturing of key components like advanced deformable mirrors and coronagraphic optics, and their amalgamation in ``adaptive coronagraph'' systems. 
Moreover, we discuss highly integrated system designs that combine contrast-enhancing techniques and characterization techniques (like high-resolution spectroscopy) while minimizing the overall complexity. 
Finally, we explore extreme implementations using all-photonics solutions for ground-based telescopes and dedicated huge apertures for space telescopes.
\end{abstract}


\section{INTRODUCTION}
This mini-review is one of the results of the ``Optimal Optical Coronagraph'' workshop held at the Lorentz Center in Leiden on September 25-29 2017, which brought together a diverse group of 25 researchers (most of the authors of this paper) to discuss and work on urgent problems in high-contrast imaging of exoplanets.
A number of interactive sessions led to new insights and collaborations.
The main outcomes of the workshop are summarized in a series of papers: Paper I on Contrast metrics and coronagraph optimization\cite{paperI}, Paper II on common-path wavefront sensing and phase retrieval techniques\cite{paperII}, and this is Paper III that reviews novel technologies for high-contrast imaging.
In this review, we will cover a broad range of optical elements, contrast-enhancing techniques, and exoplanet characterization functionalities that can be combined into systems for high-contrast imaging.
We cover both ground-based and space-based telescope implementations, which to a large extent apply the same elements. 
However, the typical time-scales on which phase aberrations vary are drastically different on the ground and in space: the Earth's atmosphere limits raw contrast on ground-based telescopes to $\sim$10$^{-3}$--$10^{-5}$ at a few diffraction widths, while a stable space observatory can dig dark holes as deep as $\sim$10$^{-11}$.
In this review, we do not split up ground-based and space-based systems, but rather  discuss the basic optical elements and the integrated optical systems for general high-contrast instruments, and highlight specific use cases for on the ground or in space.
We emphasize that this review mostly reflects the discussions during our workshop, and is therefore by no means aiming to be complete in any way.
We refer to existing reviews for more comprehensive introductions to the field of high-contrast imaging\cite{Mawetreview,Kasperreview,Millireview,GuyonELTs}.
We will discuss a number of opportunities that we identified, even though their TRL is at most 1, and the level of craziness is occasionally high.

\section{COMPONENTS FOR HIGH-CONTRAST IMAGING}

\subsection{Deformable mirrors}
Adaptive Optics (AO) forms the heart of high-contrast imaging systems, both on the ground as well as in space.
For ground-based telescopes, one or several deformable mirrors (DMs) mostly have the functionality to counter the effects of the turbulent atmosphere.
Lessons learnt from the VLT/SPHERE\cite{SPHERE} and Gemini/GPI\cite{GPI} instruments indicate that upgrades should include deformable mirrors that run at considerably higher speed; $\sim$3 kHz.
Moreover, to improve the Strehl ratio at wavelengths down to visible light (in which we want to see planets in reflected light), future DMs need more actuators (several thousand for the current 8-m class telescopes), even though their effect on raw contrast will mostly still need to be at a few $\lambda/D$ (i.e.~a few cycles over the pupil).

In space, DMs are mainly used to correct for slowly evolving small aberrations of the telescope and instrument themselves, and to dig a dark hole around the suppressed PSF core.
To enable wide and deep dark holes, both phase and amplitude aberrations need to be controlled, which is achieved by a pair of DMs, one of which is not located in the pupil and acts on amplitude through Fresnel propagation of its phase shape to the pupil plane.

\begin{figure} [t]
   \begin{center}
   \includegraphics[width=\textwidth]{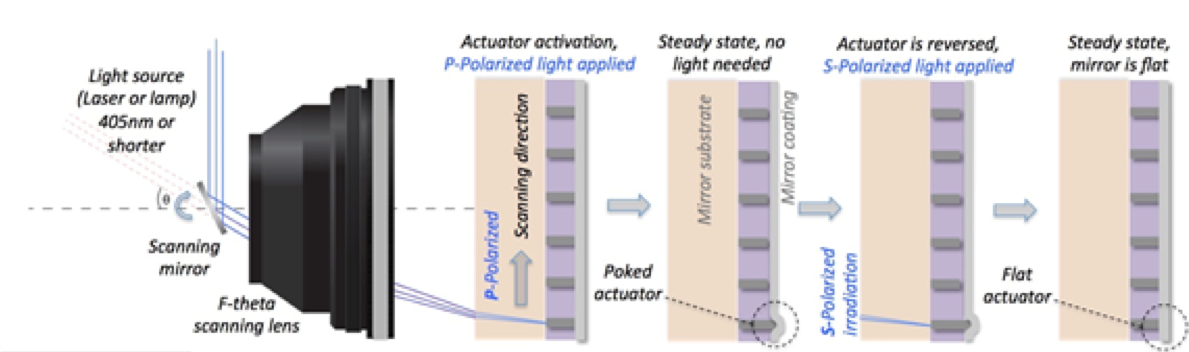}
   \end{center}
   \caption{ \label{fig:photonicDM} 
Illustration of the photonic DM concept\cite{photonicDM}. A laser source illuminates and activates the actuator, which is responsive based on the polarization state of the light.}
\end{figure}

Many different actuator types have been implemented in deformable mirrors; including piezo, voicecoil, magnetic reluctance, MEMS, bimorph, and ferrofluid actuation\cite{ferrofluidDM}.
One issue with DMs with high actuator counts is cabling.
This issue is solved by a ``photonic DM''\cite{RiaudphotonicDM,photonicDM}, which uses light to address the deformable phase-sheet to activate individual activators, see Fig.~\ref{fig:photonicDM}. 
In this implementation, a special polymer hybrid foil can be made to expand and contract by controlling direction of a laser that illuminates part of the foil.
This concept can be used to implement sparse wavefront control\cite{photonicDM} in which only a small portion of the surface is actuated to correct higher-order aberrations directly on the primary mirror or as part of a stand-alone photonic device. 
Sparse wavefront control meets the specialized needs of space-based high-contrast imaging in which faint and slowly varying speckles must be corrected at slower speeds and for which a full-surface modulation is not required.

\begin{figure} [!h]
   \begin{center}
   \includegraphics[width=0.5\textwidth]{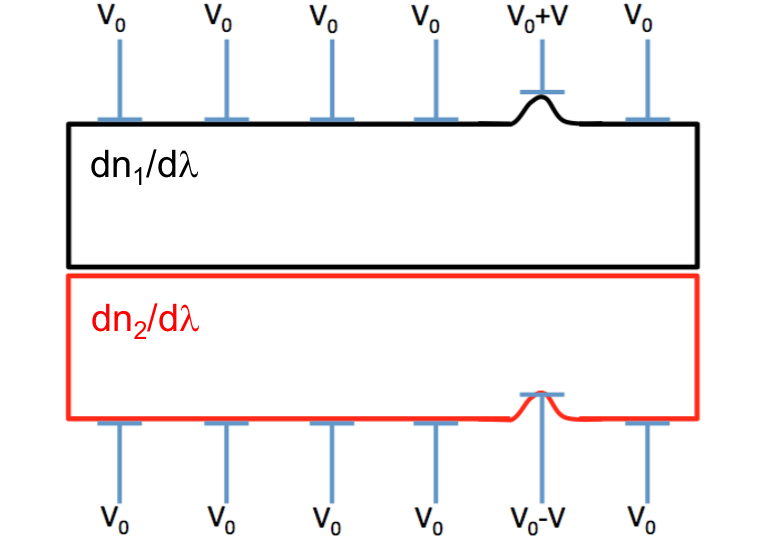}
   \end{center}
   \caption{ \label{fig:antichromatic} 
Cartoon design for an ``antichromatic'' deformable element.}
\end{figure} 

One major challenge for AO systems on the ground and particularly in space is to deliver contrast performance over a broad wavelength range, to enable spectroscopy of the target under study.
As a single deformable mirror only provides a phase shape that is independent of wavelength (in absolute sense), the resulting contrast is hence optimal at only one wavelength, as the wavefront is always to some extend chromatic due to the atmosphere and the instrument itself.
Multiple (three\cite{3DMs} or more) can be combined to deliver control over contrast over a broader spectral band, at the cost of added optomechanical and control complexity.
Adaptive optics systems can also be extended with dedicated active elements that provide absolute phase control as a function of wavelength.
Spatial Light Modulators (SLMs) constitute one possible solution, as they contain transmissive liquid-crystal pixels, which phase (including spectral dispersion; in one polarization direction) can be controlled by voltage.
SLMs generally have many pixels, and are increasingly fast.
One other possible solution, particularly for space-based high-contrast imaging, is sketched in Fig.~\ref{fig:antichromatic}.
This adaptive element is comprised of a combination of glasses with different dispersions (cf.~crown/flint of achromatic lenses).
By selectively expanding/contracting a part of each glass, e.g.~through heating using transparant electrodes, a spectral slope can be imposed on the resulting wavefront.

\subsection{Coronagraphs}
Coronagraphs are optical elements, or combinations thereof, that (locally) suppress the diffraction halo of the stellar PSF to enable direct observations of a close companion.\cite{Guyoncoro}
Coronagraphic elements can reside in the focal plane and/or pupil plane and manipulate the amplitude and/or phase of the stellar light such that diffraction rings disappear.
There is a wide range of coronagraphic techniques, and during the first Leiden workshop in 2004 a ``coronagraphic tree of life'' was constructed\cite{firstLeidenworkshop,Mawetreview}.
During the 2017 Leiden workshop it was concluded that this tree is now so convoluted, with many branches crossing and hybridizing with other coronagraphic branches and also with totally different parts of the high-contrast imaging system that it is time to cut down the tree, and discuss coronagraphic systems in all their complicated glory.
For this mini-review, we focus our intention on coronagraphic phase masks, both in the focal plane as well in the pupil plane, as they have seen the most development regarding design and manufacturing in recent years, and, moreover, they deliver the best ideal contrast performance.

\begin{figure} [ht]
   \begin{center}
   \includegraphics[height=4.7 cm]{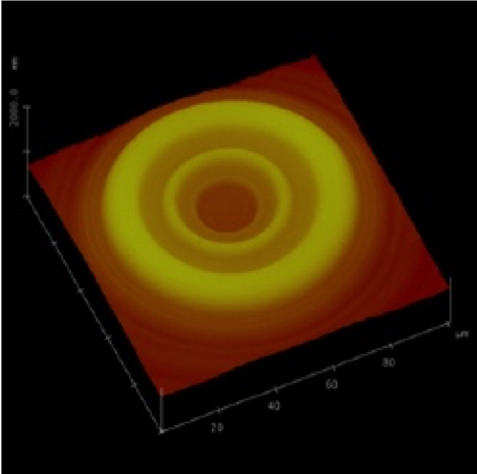}
   \includegraphics[height=4.7 cm]{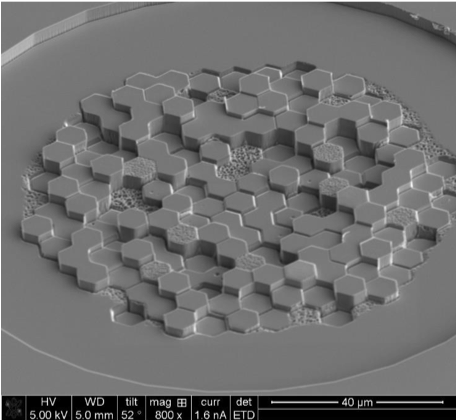}
   \includegraphics[height=4.7 cm]{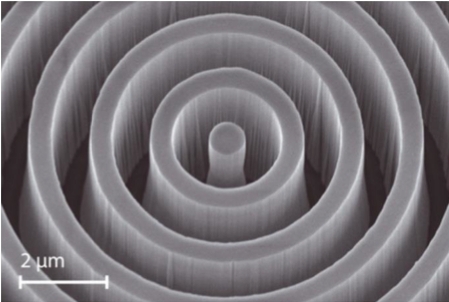}
      \end{center}
   \caption{ \label{fig:fpcoro} 
Electron microscope images of the Hybrid Lyot coronagraph mask for WFIRST\cite{hybridLyot}, the Complex Mask Coronagraph for SCExAO\cite{CMC}, and the Annular Groove Phase Mask\cite{AGPM}.}
\end{figure} 

Fig.~\ref{fig:fpcoro} displays electron microscope images of three advanced focal-plane coronagraph masks.
All three masks have been produced using lithographic techniques.
The Hybrid Lyot mask\cite{hybridLyot} consists of a metallic and a dielectric layer, patterned to deliver the desired coronagraphic phase mask, in combination with Zernike wavefront-sensing (through the $\pi$/2 dimple in the middle).
The Complex Mask Coronagraph\cite{CMC} was computer-optimized to diffract all the on-axis light outside the Subaru pupil (and inside the secondary obscuration and spiders), for a relatively broad wavelength range.
The Annular Groove Phase Mask\cite{AGPM} consists of sub-wavelength circular grooves that conspire to create a relatively broadband half-wave retarder, which fast axis forms a Vector Vortex Coronagraph that destructively interferes the on-axis light.

\begin{figure} [t]
   \begin{center}
   \includegraphics[width=\textwidth]{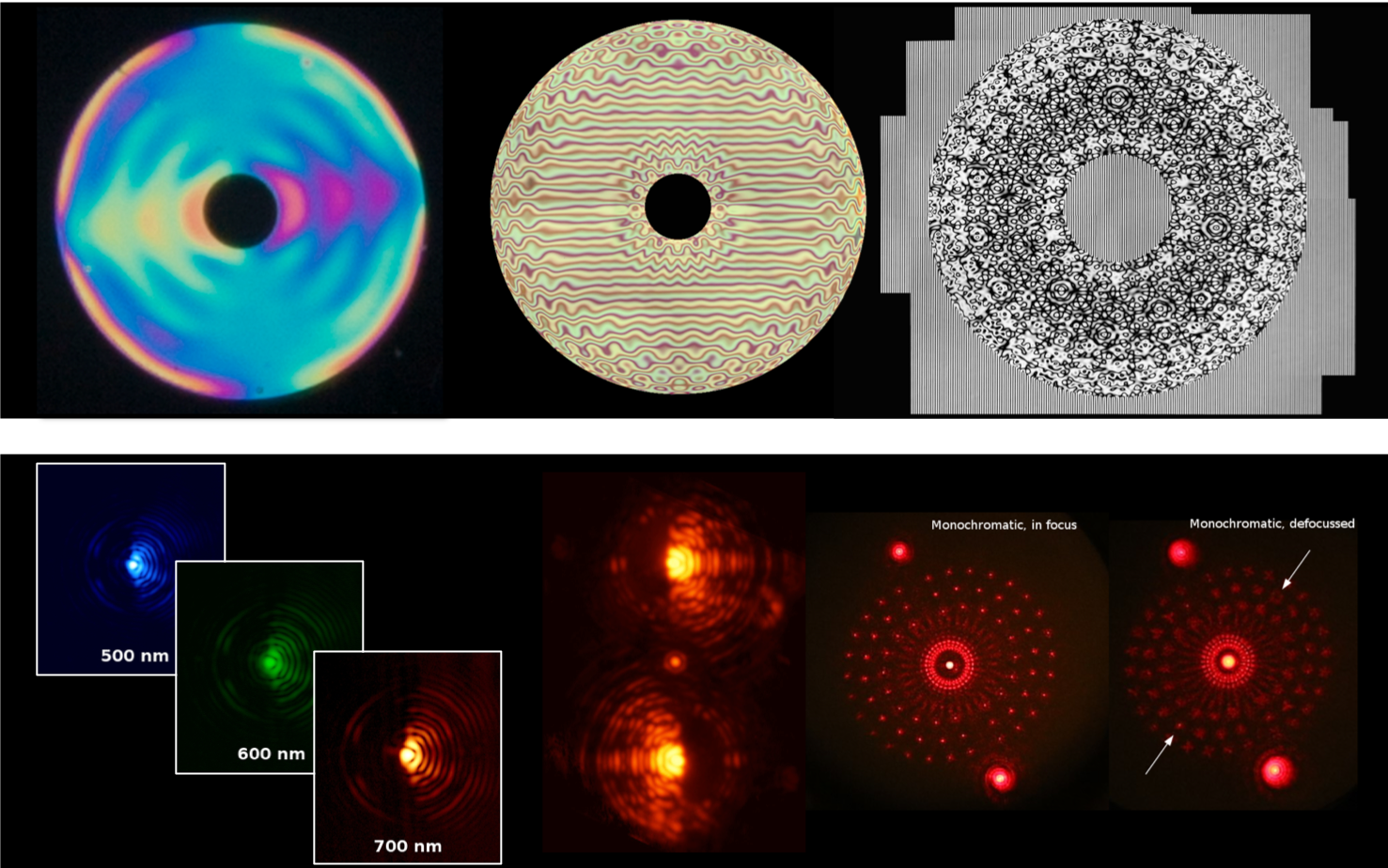}
      \end{center}
   \caption{ \label{fig:vAPP} 
\textit{Top:} stitched microscope images of vector-APP devices in between crossed polarizers\cite{OttenvAPPlab,DoelmanvAPP,WilbyHMWFS}. \textit{Bottom:} Vector-APP point-spead functions. The first and the third correspond to the pupil phase patterns above. The middle one is an on-sky result from Ref.\cite{OttenvAPPMagAO}~.}
\end{figure}

The latter is an example of a ``geometric phase'' element\cite{Mawetvortex,geometricphase} that imposes a phase pattern in a strictly achromatic fashion. When the retardance of such an element is half-wave at any given wavelength, then no leakage of the non-coronagraphic PSF occurs.
Using state-of-the-art liquid crystal techniques, geometric phase patterns can now be applied to a substrate with any possible configuration with $\sim$4 $\mu$m pixelation\cite{directwrite}, and with high diffraction efficiencies over wavelength ranges of more than an octave\cite{MTR}.
These liquid-crystal technologies are being exploited in the vector-APP coronagraph\cite{SnikvAPP}, see Fig.~\ref{fig:vAPP} for a little image gallery.
This coronagraph resides in the pupil plane and applies a (geometric) phase pattern such that the resulting PSF has a dark hole, which remains dark over a large wavelength range.
By splitting circular polarization states (for which the geometric phase imposes a phase with opposite signs), the resulting two PSFs have dark holes on either side.
As essentially \textit{any} phase pattern can be produced, the vector-APP pattern can be multiplexed with holographic spots for focal-plane wavefront sensing\cite{WilbyHMWFS}, see Fig.~\ref{fig:vAPP} on the right.

In general, technology now allows us to produce virtually any (static) phase and amplitude pattern that a computer with a good optimizer algorithm\cite{paperI} can come up with.
In addition, the combination of new technology and optimizers now opened the door for hybrid coronagraphic systems, in which pupil- and focal-plane amplitude/phase masks are combined (also with the rest of the system) to yield optimal performance.

\subsection{Detectors}
For extreme high-dynamic-range imaging, we require low-noise or no-noise detectors.
In addition, for ground-based instruments, we require very fast (several kHz) frame rates, at least for the wavefront sensor cameras, but preferentially also for the science cameras.
The SAPHIRA detector\cite{SAPHIRA} and its C-RED One incarnation offer low-noise, high-speed operation, thanks to a an avalanche photodiode array.
These are now the detectors of choice for wavefront sensing in ground-based instruments.

Microwave Kinetic Inductance Detectors (MKIDs\cite{MKIDs}) hold the promise of offering the ultimate performance.
As they are photon-counting detectors, they have no read noise, and run at the ultimate speed.
In addition, as the energy of each photon is measured, they have an intrinsic spectral resolving power ($\lambda/\Delta\lambda$$\approx$10).
They can therefore both be used for (focal-plane) wavefront sensing, as well as for integral-field spectroscopy.
They may also find a use in high-resolution spectrograph to effectuate order-sorting.
The first MKID arrays are currently being deployed at high-contrast imaging instruments\cite{MKIDs}.

\section{SYSTEMS FOR HIGH-CONTRAST IMAGING}
\begin{figure} [t]
   \begin{center}
   \includegraphics[width=0.7\textwidth]{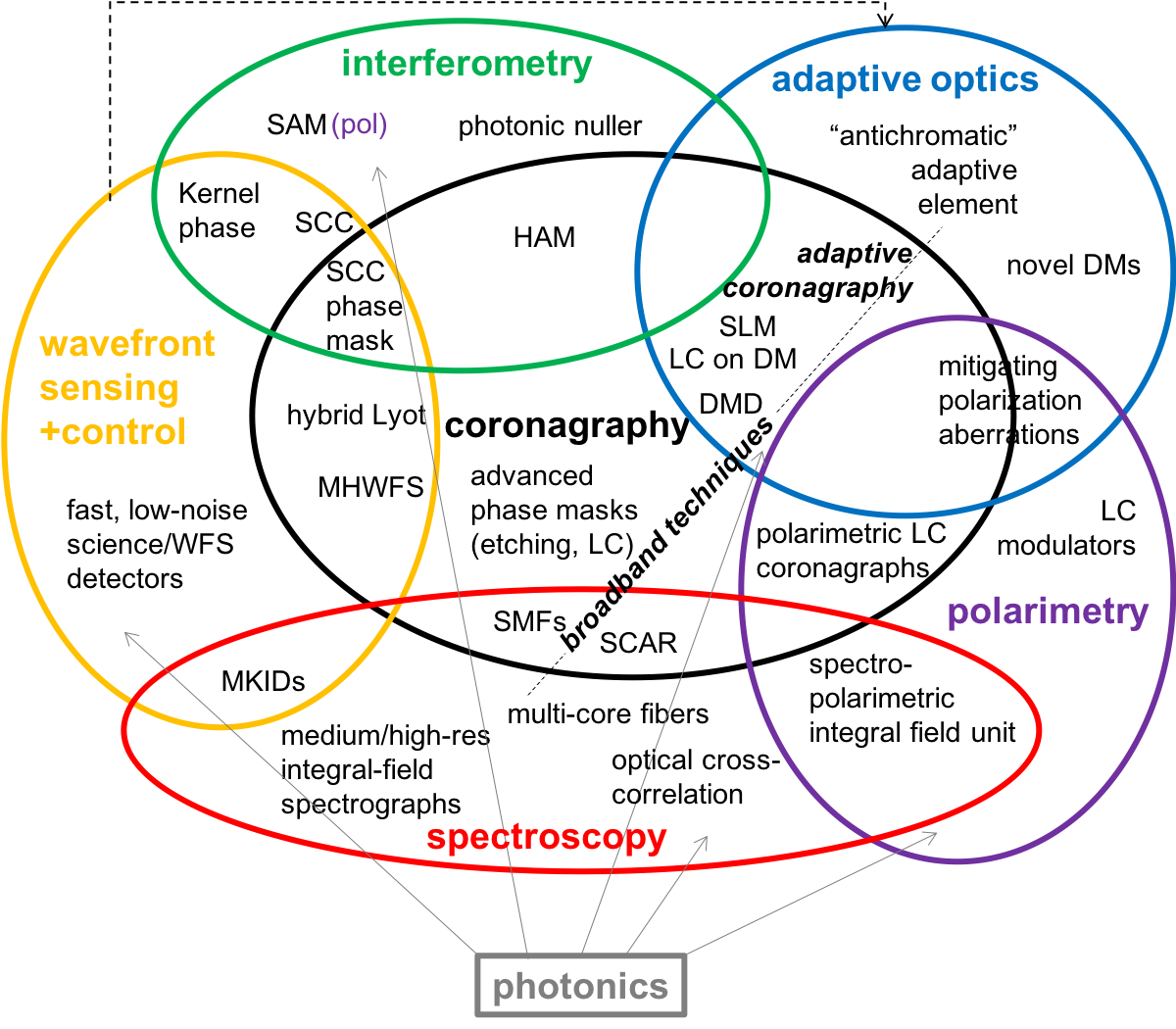}
      \end{center}
   \caption{ \label{fig:Venn} 
Two-dimensional representation of a multidimensional Venn diagram that aims to visualize the interplay between the various optical components and modalities that together constitute a high-contrast imaging system. We identify technologies that combine various functionalities in the cross-sections.}
\end{figure}

None of the optical components discussed in the previous section can achieve the required contrast performance by itself.
High-contrast imaging therefore requires, \textit{in the extreme}, a systems approach.
In Fig.~\ref{fig:Venn} we present a two-dimensional representation of a multidimensional Venn diagram that aims to visualize the interplay between the various optical components and modalities that together constitute a high-contrast imaging system.
The adaptive optics, including (focal-plane) wavefront sensing, deliver, together with the coronagraphic optics, the raw contrast.
Particularly for ground-based instruments, additional contrast-enhancing techniques like spectroscopy (including Spectral Differential Imaging and High-resolution spectroscopy) and polarimetry can deliver additional orders of magnitudes of exoplanet detectability in the residual stellar speckle halo.
At the same time, the furnish the diagnostic capability to \textit{characterize} the atmospheres and possible surfaces of the exoplanets under observation.

The complicated Venn diagram of Fig.~\ref{fig:Venn} mostly serves to identify technologies that combine different functionalities for high-contrast imaging, and can thus facilitate systems-level design choices for the increasingly integrated instruments of the future.

\subsection{Adaptive Coronagraphy}
One obvious merger of high-contrast imaging functionalities, enabled by new technologies, is the amalgamation of adaptive optics and coronagraphy into ``adaptive coronagraphy''.
Both separate techniques aim to shape the stellar PSF such that companions can most optimally be detected at a small angular separation from the star.
Increasingly, coronagraphy has occupied the pupil plane to effectuate this beam-shaping.
Therefore, a tandem functionality in the pupil plane between correcting wavefront errors and ``PSF engineering'' is an obvious direction.
Note that the joint goal of adaptive coronagraphy is not to create a flat wavefront; the goal is to achieve contrast through whatever means.
An adaptive coronagraph system has many benefits, as the system can be adapted to both the observational circumstances, and the actual science case.
As for the environmental factors, such a system can
\begin{itemize}
\item adapt to the state of the telescope (e.g.~missing segments);
\item adapt to the observing conditions (good/bad seeing);
\item adaptively correct optical errors that degrade contrast.
\end{itemize}
And as for the scientific goals, such a system can
\begin{itemize}
\item create a large dark hole to enable direct detection of exoplanets;
\item create a deep, small and co-rotating dark hole at a known position of an exoplanet, to enable characterization;
\item accomplish contrast performance around two (or more) binary (or multiple) stars in the field of view.
\end{itemize}

A first implementation of adaptive coronagraphy is through leveraging of the DM(s) to dig coronagraphic dark holes\cite{paperII}.
The most generic formulation for any telescope aperture is through the Active Correction of Aperture Discontinuities\cite{ACAD}, which uses the common two-DM architecture of space-based high-contrast systems to augment the coronagraphic performance.
Novel optical elements that derive from the display/projector industry are currently being successfully tested for adaptive coronagraphy with a large number of degrees of freedom (i.e.~programmable pixels).
Spatial Light Modulators (SLMs) can be used to implement complicated phase patterns in the focal plane and/or the pupil in quasi-real-time\cite{SLM}.
And Digital Mirror Devices (DMDs) can be used for adaptive amplitude apodization, currently in binary black/white mode\cite{DMD}, but possibly extended to greyscale operation through time modulation.

\subsection{Spectroscopy}\label{sect:spectroscopy}
Spectroscopy is an obvious ``back-end'' technique for high-contrast imaging to provide characterization diagnostics of targets under study.
In addition, it also provides significant contrast-enhancement, particularly for ground-based instruments.
Most current high-contrast imagers have implemented a low-resolution Integral Field Spectrograph that enables distinguishing the faint signal of an exoplanet at a fixed angular separation from the residual speckle pattern that grows with wavelength\cite{SparksFord}.
Such IFS data also deliver low-resolution spectra of exoplanets for first-order characterization.
However, it requires medium spectral resolution to really discern the instrumental effects as a function of wavelength\cite{Konopacky}.

Even though direct observations of faint exoplanets are starved for useful photons from the target, at has recently been shown\cite{Snellen2015} that splitting up the light in high-resolution spectra is beneficial, as most spectral lines (in reflection or emission from the planet) that can be distinguished from direct stellar lines only become sharp and deep features at high spectral resolution.
Individual spectral lines from the planet typical drown in the photon noise from the residual stellar speckles, but a cross-correlation of the spectral data with a template consisting of a forest of spectral lines at known wavelengths, originating from known atmospheric constituents (e.g.~CO, H$_2$O, O$_2$), provides both the necessary contrast enhancement and the characterization of the exoplanet at the same time.
We therefore require spectrographs or rather spectroscopic integral-field units at medium and high spectral resolution to achieve both the best contrast performance, and the full range of spectral diagnostics.

\begin{figure} [t]
   \begin{center}
   \includegraphics[width=0.9\textwidth]{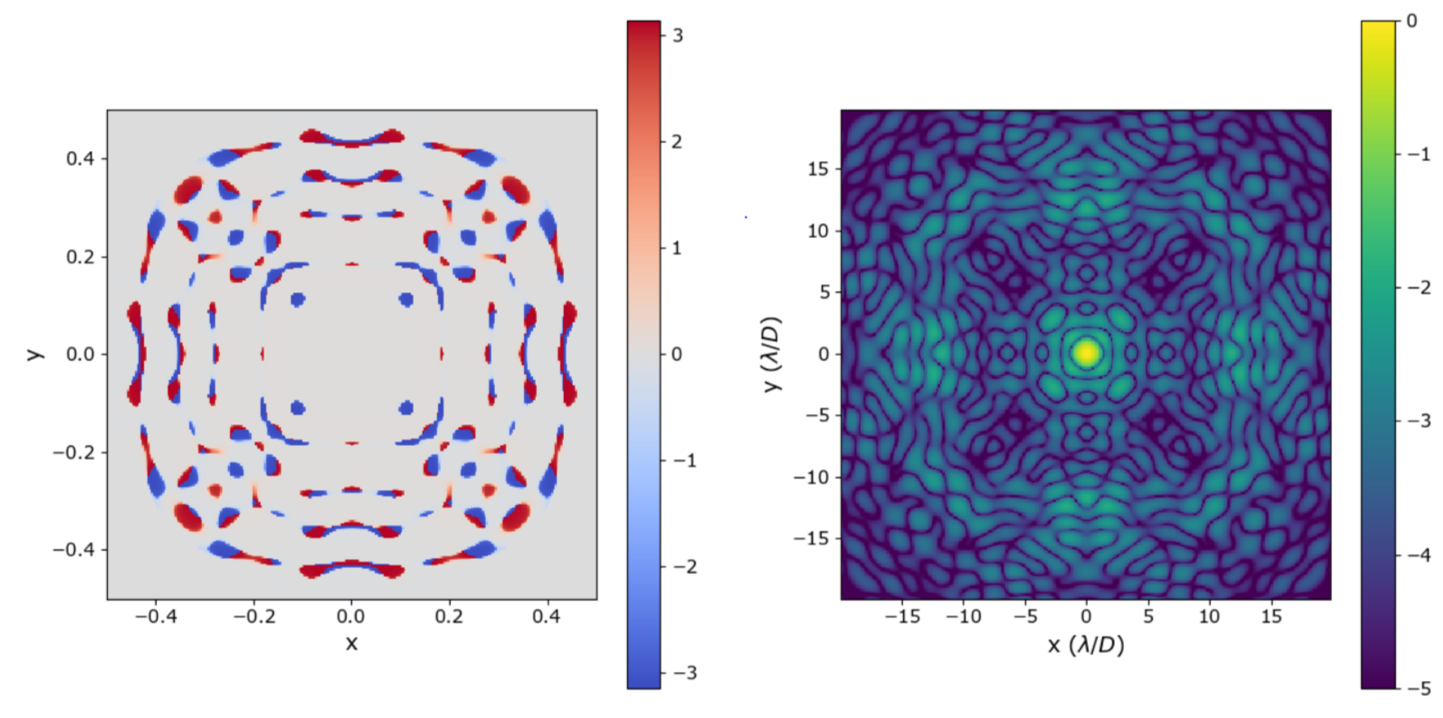}\\
   \includegraphics[height=6.5 cm]{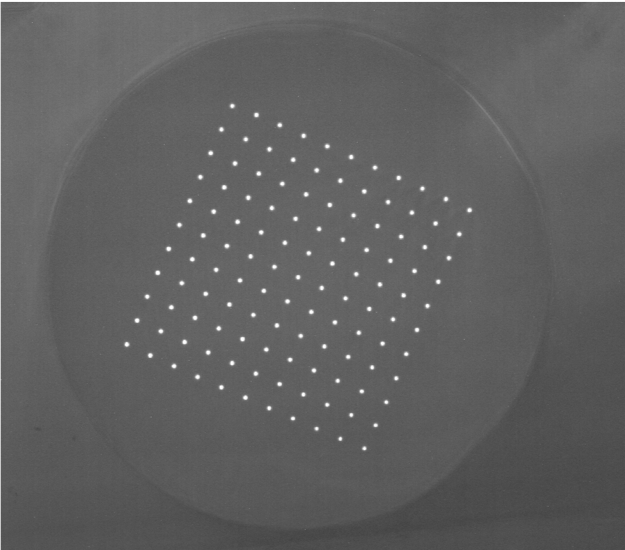}
   \includegraphics[height=6.5 cm]{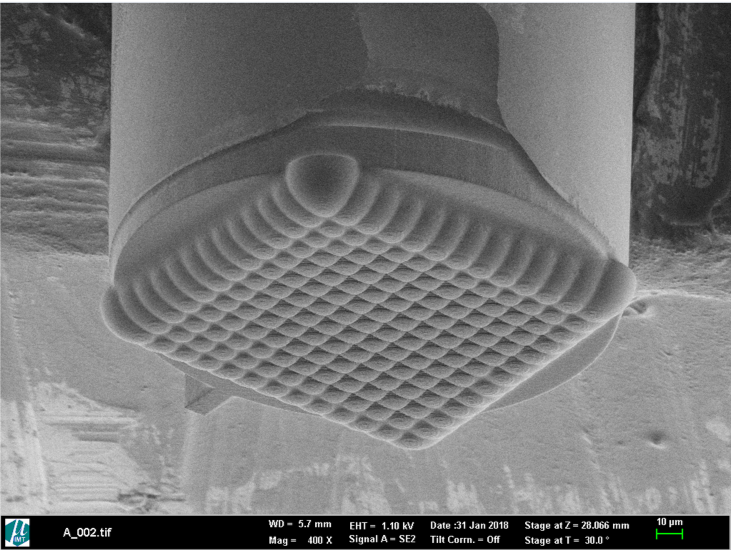}
      \end{center}
   \caption{ \label{fig:multicoreSCAR} 
Example of a SCAR coronagraph system. The pupil phase pattern (\textit{top left}) creates a PSF (\textit{top right}) that contains two nulls for every single-mode fiber in the multi-core fiber (\textit{bottom left}) when fed through the 3D-printed microlens array (\textit{bottom right}). [\textit{S.~Haffert\cite{SCARIFU1}, E.~Por\cite{SCARIFU2}, M.~Blaicher}]}
\end{figure}

A crucial enabling technology to combine high-contrast imaging and high-resolution spectroscopy has been found in the form of single-mode fibers (SMF)\cite{fiberinjection,SMF, SCAR1}. 
Such single-mode fibers do not only serve as a conduits of light from a high-contrast focal plane; they also act as natural coronagraphs as they only accept the fundamental mode which roughly takes the shape of an on-axis PSF. 
As such, a SMF is naturally matched to feed the diffraction-limited core of an exoplanet image.
Moreover, it naturally reject off-axis light from the stellar PSF, particularly if the electric field goes through one or two nulls over the mode-field diameter.
This property is exploited by the ``SCAR'' coronagraph system\cite{SCAR1,SCAR2} that shapes the off-axis stellar PSF (with a pupil-plane phase pattern \`a la the vector-APP) to minimize the off-axis coupling into single-mode fibers through a microlens array, see Fig.~\ref{fig:multicoreSCAR}.
This means that the PSF itself does not necessarily have dark regions anymore, but that the entire end-to-end system is optimized for contrast performance (including tip/tilt sensitivity and broadband use).

In addition to patterned liquid-crystal technologies, SCAR can profit from brand-new technologies.
An integral-field unit can now be constructed from multi-core fibers\cite{multicore} that combine many single-mode fibers in a single cladding.
The microlens array that feeds every individual fiber can now be 3D-printed\cite{nanoprinting} on top of this multi-core fiber, see Fig.~\ref{fig:multicoreSCAR}.
The nanoprinting has sufficient resolution that even compound or aspheric lenses (e.g.~PIAA) can be manufactured\cite{nanoprinting}.
The resulting compound fiber-head is very easy to install in a focal plane of a high-contrast imaging instrument.
In addition, the use of single-mode fibers or indeed multi-single-mode-fiber cores has the additional benefit that the corresponding (integral-field) spectrograph can be extremely compact\cite{LEXIspectrograph}.

One disadvantage of high-resolution spectroscopy, particularly when combined with an integral-field unit, is that it requires detectors with copious amounts of pixels to sample the spectra.
This situation can be alleviated by optical cross-correlation techniques, that would permit the use of e.g.~a lower-resolution spectrograph or a much higher degree of spatial multiplexing.
One example comprises interferometric heterodyning of quasi-periodic spectral signatures of molecules down to a lower spectral resolution\cite{heterodyne}.
An ultimate system would produce cross-correlation spectra instead of high-resolution spectra for every ``spaxel'' in the field of view, with functionality to program any cross-correlation template as desired.

\begin{figure} [t]
   \begin{center}
   \includegraphics[width=0.9\textwidth]{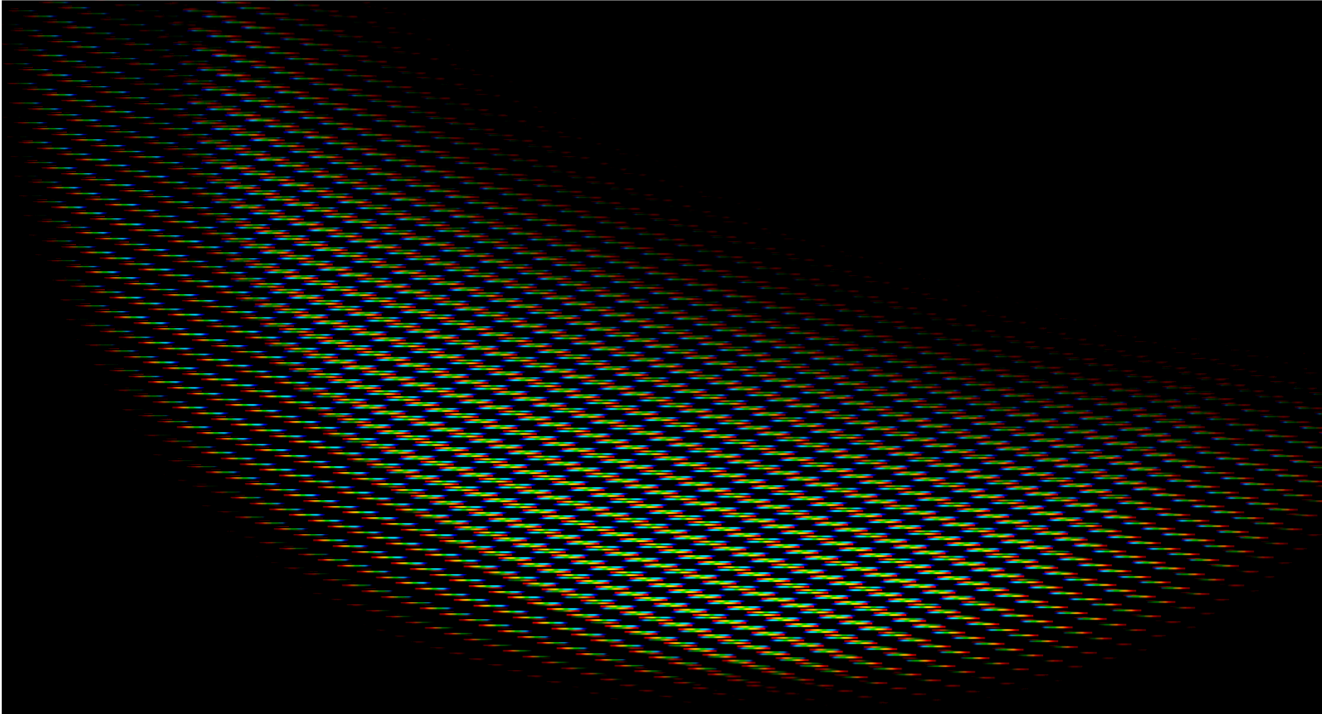}
      \end{center}
   \caption{ \label{fig:polIFU} 
Observation of Venus with a spectropolarimetric integral-field unit. For every spaxel both the spectral and the polarimetric information is retrieved in one shot.\cite{polIFU}}
\end{figure} 

\subsection{Polarimetry}
Like spectroscopy, polarimetry can deliver both contrast-enhancement and characterization functionality.
Polarimetric modes at current ground-based high-contrast imagers are very scientifically productive for detecting, imaging, and charactering protoplanetary disks in scattered light.
For the ultimate goal of observing an Earth-like planet in reflected light, polarimetry will be crucial to discern residual unpolarized starlight from polarized light from the exoplanet.
In fact, to detect the presence of O$_2$ on a life-bearing planet, one needs to distinguish its signal from that of our own O$_2$ (in the case of ground-based observations with Extremely Large Telescopes), and the two distinct properties are that an exoplanet's O$_2$ is generally Doppler-shifted (spectral characteristic) and it is polarized.
In fact, generally only the combination of spectroscopy and polarimetry in spectropolarimetry can provide unambiguous characterization of exoplanetary atmospheres.
Moreover, the polarization angle of a detected source immediately confirms companionship.

Modern polarimetric techniques were recently reviewed in Refs.\cite{Snikpolreview1,Snikpolreview2}~.
In this discussion about high-contrast imaging systems, we highlight an elegant solution to implement a spectropolarimetric integral-field unit\cite{polIFU}, see Fig.~\ref{fig:polIFU}.
This implementatation is enabled by a ``polarization grating''\cite{polarizationgrating}, a liquid-crystal element that furnishes very efficient diffraction in orders $\pm$1 over, e.g., the entire visible wavelength range, whilst splitting opposite circular polarization states into the two diffraction orders.
By adding an achromatic quarter-wave plate and a polarization modulator (e.g.~rotating half-wave plate or switching liquid crystal), this unit measures linear polarization as a function of wavelength for every pixel in the field of view, as sampled by the microlens array.
As both polarization states are sampled in the same way, there are no differential aberrations to degrade the polarimetric performance.

Note that polarization optics are also required to mitigate the effects of polarization aberrations\cite{polarizationaberrations}, that broaden and distort the PSF, especially for extremely high-contrast observations in space.

\begin{figure} [p]
   \begin{center}
   \includegraphics[width=0.8\textwidth]{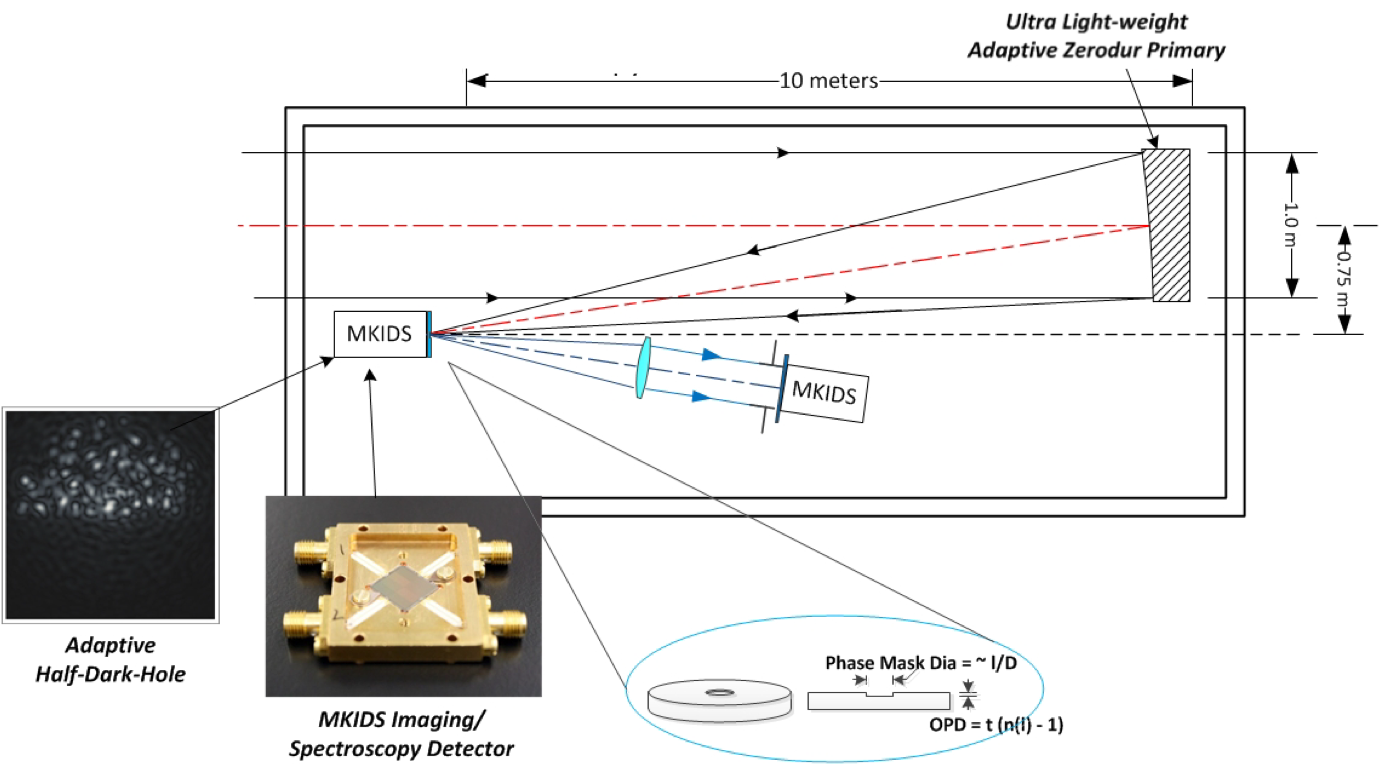}\\
   \vspace{5mm}
   \includegraphics[width=0.8\textwidth]{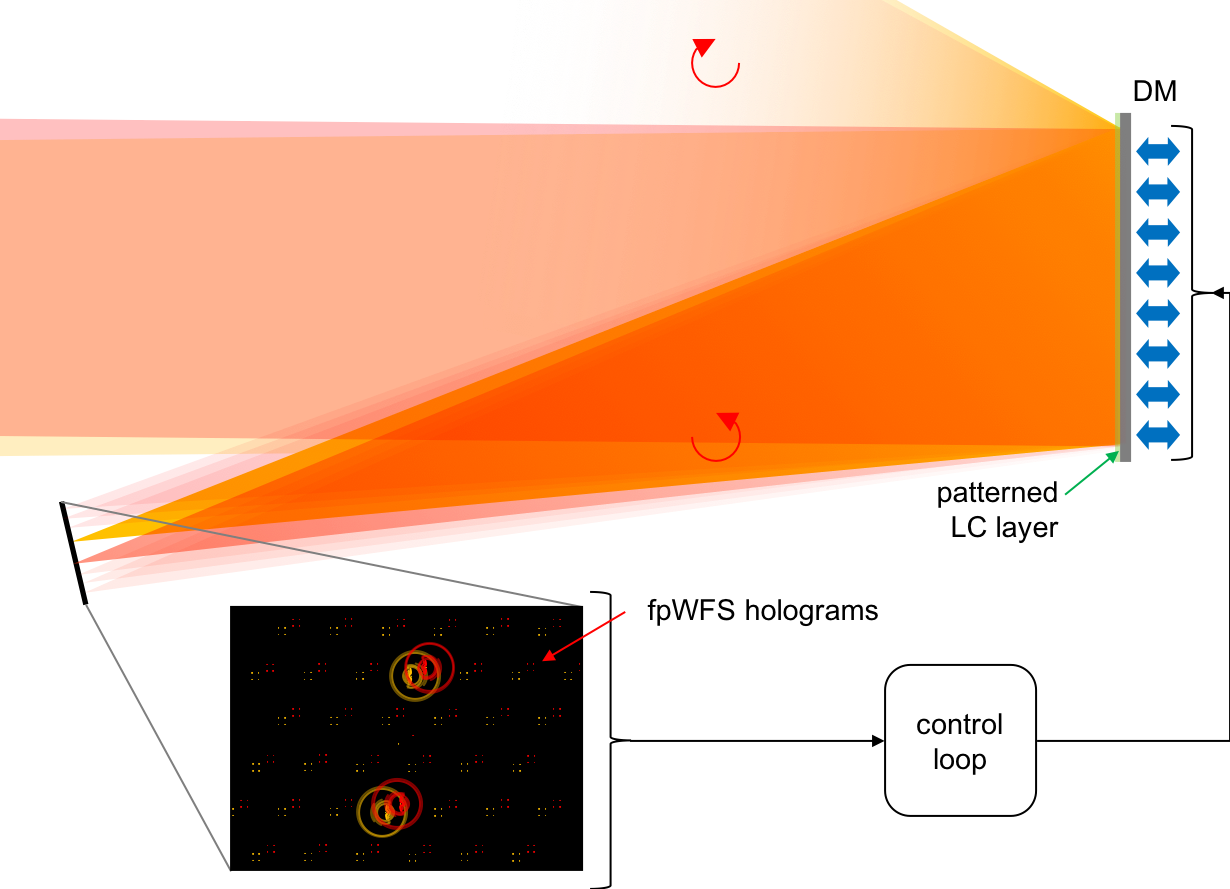}
      \end{center}
   \caption{ \label{fig:minimaltelescope} 
   Two examples of a ``minimal instrument design'' for a space telescope. \textit{Top:} An adaptive powered primary mirror focuses the light onto a focal-plane mask on top of an MKID imaging detector. The mask acts both as the coronagraph as well as a Zernike wavefront sensor, which light is analyzed by a second MKID. The MKIDs offer no-noise photon-counting spectrally resolved imaging. [\textit{J.~K.~Wallace}] \textit{Bottom:} Concept sketch for a $\sim$30-cm space telescope to image the habitable zones of $\alpha$ Cen A and B. The primary is adaptive and flat, and is covered by patterned liquid crystals that focus the light, impose coronagraphic dark holes for both stars regardless of roll angle, implement focal-plane wavefront sensing spots, and enable spectroscopy. 
   [\textit{F.~Snik}]}
\end{figure} 

\subsection{``Minimal Systems''}
It is an interesting, perhaps somewhat academic exercise, to design high-contrast imaging systems with a minimal amount of components, which therefore have to combine several functionalities.
One goal is to minimize the overall system complexity, and optimize the performance for one scientific target (high-contrast observations of exoplanets or one particular subset of exoplanets), and that scientific target only.
For space-based systems, a highly compact design will greatly reduce the launch cost, and therefore the viability of the mission.
For instance, to perform direct imaging of the habitable zones of our neighboring stars $\alpha$ Cen A and B, one only needs a space telescope with a $\sim$30-cm aperture, operating in visible light\cite{ACEsat}.

Fig.~\ref{fig:minimaltelescope} introduces two concepts for space telescopes which both consist of a single (primary, deformable) mirror and an imaging detector.
In the first case, the coronagraphy, wavefront sensing, and spectroscopy is integrated on top of the imaging detector, an MKID.
A second MKID analyzes the light reflected by the focal-plane Zernike mask, and provides a second layer of wavefront sensing.
In the second case, the deformable primary is flat, and covered by patterned liquid crystals that perform several tasks at once: focusing the light (for one circular polarization), imposing dark holes for both binary stars regardless of roll angle, and adding focal-plane wavefront sensing spots based on which the deformable primary can be controlled.
As this is a diffractive telescope, the instantaneous bandwidth is necessarily very narrow, but space accommodates long exposure times.
Furthermore, as the diffraction angle varies with wavelength, several images in different wavelengths can be accumulated in parallel on one or several detectors equipped with, e.g., a stepped filter and microlenses to reformat the beam.

One major benefit of such systems is that the telescope pupil, the adaptive optics, and the coronagraph coincide, which removes the need for optical relays and two-DM systems for phase+amplitude correction.
Also, for the second system pointing is not critical.
However, in the end it is always a trade-off between system complexity and technology immaturity.

\subsection{Huge Space Telescopes}
In the end, it is clear that we need extremely large telescopes in space to be able to take images and spectra of many habitable-zone planets orbiting nearby stars.
A large swath of wild ideas have been put forward to build such telescopes in the not-so-distant future, ranging from self-assembling megastructures to metal-painted space balloons, from large grating apertures\cite{Dittoscope} to ``orbiting rainbows'' of diffracting particles\cite{orbitingrainbow}, and even using the sun as a gravitational lens\cite{sunlens}.
We indulged ourselves in similar thought experiments, and focused on two relatively viable concepts: a large spinning one-dimensional aperture (``windmill telescope''), and foldable large diffractive apertures.

\begin{figure} [!h]
   \begin{center}
   \includegraphics[width=0.7\textwidth]{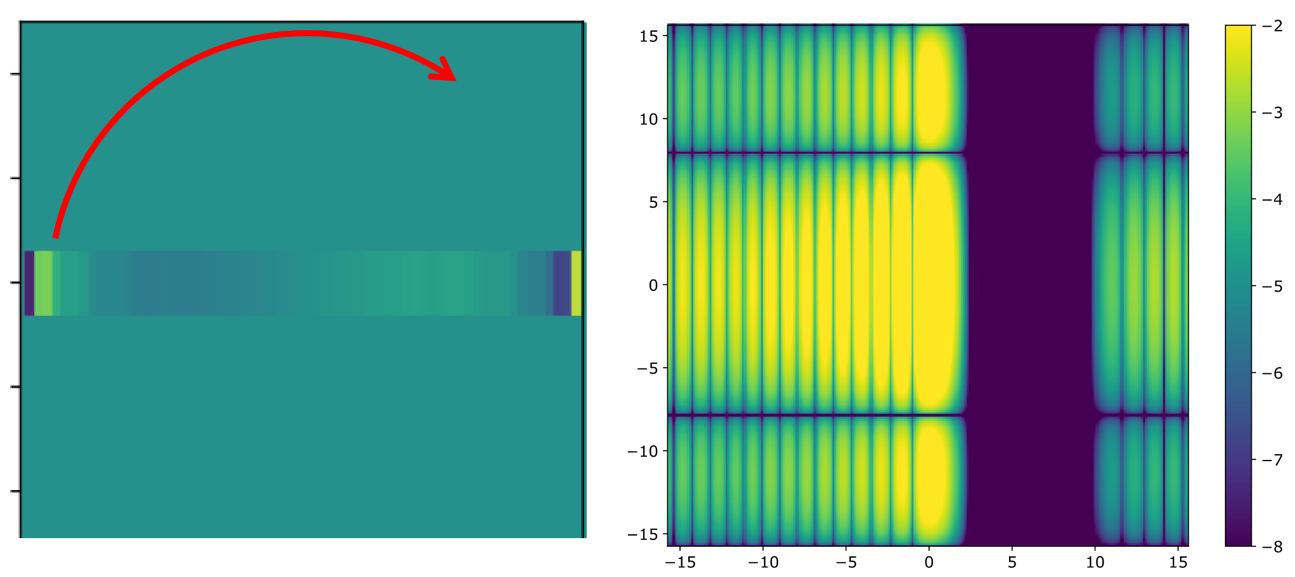}
      \end{center}
   \caption{ \label{fig:windmill} 
A windmill telescope with a vector-APP-like phase pattern on top to create a dark hole in the direction of highest spatial resolution. [\textit{E.~Por}]}
\end{figure}

Such ``windmill'' apertures have already been discussed in the literature\cite{windmill1, windmill2}.
Their main advantage is that they offer the largest-possible aperture in one direction, given the constraints of a launch vehicle.
By rotating the aperture, a synthetic PSF can be built up.
To create a dark hole, Fig.~\ref{fig:windmill} introduces an additional vector-APP-like phase pattern on top of the aperture, which could be mosaicked together from individual patterned liquid-crystal patches, or by deforming the mirror in this particular fashion.

An even larger aperture telescope can be fit in a launch vehicle if it can be folded.
One promising technique is derived from Fresnel zone plates\cite{Fresnellens}, which turns the annular patterns into hole patterns that can be manufactured in polymer foils\cite{photonsieve1}.
Such ``photon sieve'' designs can be augmented much in the same way as shaped pupils to modify the emerging PSF to exhibit dark holes.
As this is a diffractive telescope, its use is classically restricted to narrow-band use, but combinations with regular phase elements can extend the bandwidth significantly\cite{photonsieve2}.
We already know from starshade prototyping that very large space foils can be folded up into quite a small package\cite{starshade}.
It is intriguing to hypothesize how the different polymer technologies discussed in this review can be combined to form a multifunctional foldable large-aperture space telescope: the photon sieve holes for diffractive imaging and shaped-pupil-like coronagraphy, patterned liquid crystals for extending the spectral band and implementing vector-APP-like coronagraphy, and the photonic DM materials for active shape control using a polarization-modulated laser from the (separate?) spacecraft body.

\subsection{Integrated Photonics Solutions}
In many areas of astronomical instrumentation the functionalities of classical optical components are being taken over by miniaturized photonic components, and in the next few years this will also be the case for high-contrast imaging.
The largest success stories of ``astrophotonics'' pertain to interferometry, both for multi-telescope beam combination, as well as for Sparse Aperture Masking implementations with several fibers spread in a non-redundant fashion over the pupil, which are consequently pair-wise combined and spectrally dispersed\cite{Norrisreview, FIRST}.

\begin{figure} [ht]
   \begin{center}
   \includegraphics[width=\textwidth]{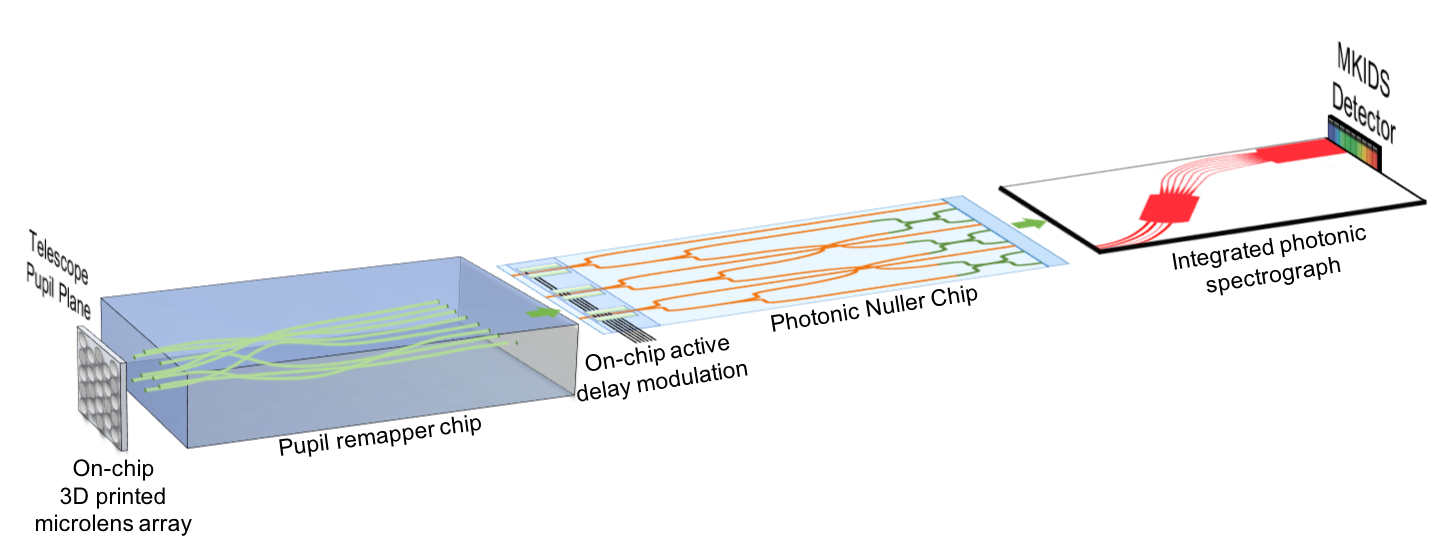}
      \end{center}
   \caption{ \label{fig:photonics} 
Conceptual sketch for an ultimate photonic instrument for high-contrast imaging. [\textit{N. Jovanovic, B. Norris, N. Cvetojevic}]}
\end{figure} 

Fig.~\ref{fig:photonics} sketches the design of an `ultimate' photonic instrument for high-contrast imaging, which includes nulling interferometry and spectroscopy in a combination of photonic building blocks.
First a 3D waveguide produced by ultrafast laser inscription in a block of non-linear glass\cite{3Dwaveguide} reformats the pupil into single-mode fiber conduits.
Injection is enabled by a 3D-printed microlens array on top of the glas.
The pupil subapertures are then pairwise interferometrically combined and nulled in half the channels.
The null can be achieved by modulating a piezoelectric material like LiNbO$_3$\cite{nulling+modulating} taking on the role of a deformable mirror, or by implementing patterned liquid crystals as achromatic nullers or active modulators.
All individual channels can then be spectroscopically analysed by on-chip arrayed waveguide spectrometers\cite{photonicspectrograph}, feeding a detector like an MKIDs.

For non-interferometric systems, a perfect coronagraph operator can also be represented by photonic combiner blocks\cite{Jewellcoro}.
Moreover, for single-mode fiber injection at the focal plane (see Sect.~\ref{sect:spectroscopy}) can be expanded with fibers that carry light with information about the low-order wavefront\cite{fiberWFS}.
With these building blocks, one day, most if not all elements of a high-contrast imaging system can be built in ultracompact integrated photonic device.

\newpage
\section{CONCLUSIONS}
\begin{itemize}
\item Instrumentation for high-contrast imaging is profiting tremendously from technology invented for completely different purposes: telecommunication, liquid-crystal displays, lithography, projectors, 3D-printing, etc.
\item However, astronomers need to adopt and appropriate these technologies to develop them in ways to deliver the optimal performances for high-contrast imaging.
\item In the end, it is only the complete end-to-end system performance that matters.
\item It is essential to try out new technologies at a very early stage on-sky, e.g.~at SCExAO\cite{SCExAO}.
\end{itemize}

We aim to expand this review and submit it as part of a series of publications to JATIS. 
We will then also update all the references to 2018 SPIE papers.

\acknowledgments 
The authors would like to acknowledge the Lorentz Center for hosting and to a large extend funding the Optimal Optical Coronagraph workshop in Leiden 2017. Additional funding for the workshop was provided by the European Research Council under ERC Starting Grant agreement 678194 (FALCONER) granted to Frans Snik. This formed the platform where this work was carried out. 

\scriptsize{
\bibliography{OOCpaperIII} 
\bibliographystyle{spiebib} 
}

\end{document}